\documentclass[aps,onecolumn,superscriptaddress,groupedaddress,nofootinbib]{revtex4}
\usepackage{graphicx}
\usepackage{dcolumn}
\usepackage{bm}
\usepackage{mathrsfs}
\usepackage{hyperref}
\usepackage{amsmath}
\usepackage{amssymb}
\usepackage{bm}
\usepackage{color}
\usepackage{float}
\usepackage{dcolumn}
\usepackage{multirow}
\usepackage{changepage}
\usepackage{enumerate}
\hypersetup{pdftex,colorlinks=true,linkcolor=blue,citecolor=red,menucolor=black,urlcolor=blue,filecolor=blue}

\newcommand{\fm}{\textrm{fm}}
\newcommand{\mev}{\textrm{ MeV}}
\newcommand{\kev}{\textrm{ keV}}

\begin{document}
\title{The isospin and compositeness of the $T_{cc}(3875)$ state }

\author{L. R. Dai}
\email[]{dailianrong@zjhu.edu.cn}
\affiliation{School of Science, Huzhou University, Huzhou 313000, Zhejiang, China}
\affiliation{Departamento de F\'{\i}sica Te\'orica and IFIC, Centro Mixto Universidad de Valencia-CSIC Institutos de Investigaci\'on de Paterna, Aptdo.22085, 46071 Valencia, Spain}

\author{L. M. Abreu}
\email[]{luciano.abreu@ufba.br}
\affiliation{Instituto de F\'{\i}sica, Universidade Federal da Bahia, Campus Universit\'{a}rio de Ondina, 40170-115 Bahia, Brazil}
\affiliation{Departamento de F\'{\i}sica Te\'orica and IFIC, Centro Mixto Universidad de Valencia-CSIC Institutos de Investigaci\'on de Paterna, Aptdo.22085, 46071 Valencia, Spain}

\author{A. Feijoo}
\email[]{edfeijoo@ific.uv.es}
\affiliation{Departamento de F\'{\i}sica Te\'orica and IFIC, Centro Mixto Universidad de Valencia-CSIC Institutos de Investigaci\'on de Paterna, Aptdo.22085, 46071 Valencia, Spain}

\author{E. Oset}
\email[]{oset@ific.uv.es}
\affiliation{Departamento de F\'{\i}sica Te\'orica and IFIC, Centro Mixto Universidad de Valencia-CSIC Institutos de Investigaci\'on de Paterna, Aptdo.22085, 46071 Valencia, Spain}

\begin{abstract}
We perform a fit to the LHCb data on the $T_{cc}(3875)$ state in order to determine its nature. We use a general framework that allows to have
the  $D^0 D^{*+}$, $D^+ D^{*0}$ components forming a molecular state, as well as a possible nonmolecular state or contributions from missing coupled channels. From the fits to the data we conclude that the state observed is clearly of molecular nature from the  $D^0 D^{*+}$, $D^+ D^{*0}$  components and the possible contribution of a nonmolecular state or missing channels is smaller than 3\%, compatible with zero. We also determine that the state has isospin $I=0$ with a minor isospin breaking from the different masses of the channels involved, and the probabilities of the  $D^0 D^{*+}$, $D^+ D^{*0}$ channels are of the order of 69\% and 29\% with uncertainties of 1\%. The differences between these probabilities should not be interpreted as a measure of the isospin violation. Due to the short range of the strong interaction where the isospin is manifested, the isospin nature is provided by the couplings of the state found to the  $D^0 D^{*+}$, $D^+ D^{*0}$ components, and our results for these couplings indicate that we have an $I=0$ state with a very small isospin breaking. We also find that the potential obtained provides a repulsive interaction in $I=1$, preventing the formation of an $I=1$ state, in agreement with what is observed in the experiment.
\end{abstract}

\maketitle

\section{Introduction}

The discovery of the $T_{cc}(3875)$ reported in \cite{expe1,expe2} was a turning point in hadron physics, showing the first evidence of a meson state clearly exotic with two open charm quarks. The state, showing as a very narrow peak in the $D^0 D^0 \pi^+$ mass distribution very close to the $D^0 D^{*+}$ and $D^+ D^{*0}$ thresholds, has spurred lots of work in the hadron community, trying to find out its origin and nature. It should be quoted that prior to this discovery there were many works looking for tetraquark structures which predicted the existence of a state of this nature with two heavy quarks \cite{6,7,8,9,10,11,12,13,14,15,17,18,19,20,22}, yet with a large span in the mass from about 250 MeV below to about 250 MeV above the actual $T_{cc}(3875)$ mass.  There were also works predicting such an object as a molecular state stemming from the interaction of the $D^0 D^{*+}$ and $D^+D^{*0}$ pairs in \cite{16,21,24}. Actually, the proximity of the mass of the  $T_{cc}(3875)$ to the $D^0 D^{*+}$ and $D^+ D^{*0}$ thresholds will guarantee that a molecule is made provided a reasonable attractive force between the pseudoscalar and vector-meson components appears \cite{25}. The existence of such a state could also be concluded from the results of the $D^* D^*$ interaction in \cite{26}, where an extrapolation of the local hidden gauge approach for meson-meson interaction \cite{hid1,hid2,hid3,hid} was used and sufficient attraction was found to form a bound state. The extreme proximity of the $T_{cc}(3875)$ state to the $D^0 D^{*+}$ threshold, of about 360 keV, has motivated many works to explain this state as a molecular state of the $D^0 D^{*+}$ and $D^+ D^{*0}$ channels \cite{Ling:2021bir,Dong:2021bvy,Xin:2021wcr,Huang:2021urd,Fleming:2021wmk,Ren:2021dsi,migueltcc,juanguo,Deng:2021gnb,Zhao:2021cvg,Ke:2021rxd,Agaev:2022ast,He:2022rta,Padmanath:2022cvl,migueljuan,Cheng:2022qcm,Abreu:2022sra,Chen:2022vpo,Jia:2022qwr,Wang:2022jop,Dai:2023mxm,Li:2023hpk,Ozdem:2021hmk,Chen:2021cfl}. The compact tetraquark structure has also had its supporters in \cite{Agaev:2021vur,Jin:2021cxj,Wu:2022gie,Meng:2022ozq}. Some works suggest that there should be a molecular state with a small admixture of a compact tetraquark \cite{Yan:2021wdl}. A comparative study of the tetraquark and  molecular structures is done in \cite{Chen:2021tnn} showing that the compact tetraquarks are about 100 MeV more bound than the molecular ones. Other works suggest to measure magnetic moments to elucidate the nature of the states \cite{Ozdem:2021hmk,Azizi:2021aib}. As mentioned, the proximity of the mass of the  $T_{cc}(3875)$
state to the $D^0 D^{*+}$, $D^+ D^{*0}$ thresholds, makes the molecular picture very natural and this can explain the large amount of works supporting this picture, mostly by using effective meson-meson potentials, or quark model interaction \cite{Chen:2021cfl}, QCD sum rules \cite{Ozdem:2021hmk,Agaev:2022ast} or lattice QCD simulations  \cite{Padmanath:2022cvl,Chen:2022vpo}. Further discussions concerning the $T_{cc}(3875)$ and related states can be seen in \cite{Meng:2022ozq,Chen:2022asf,Weng:2021hje,add1,add2}.

  While the proximity of the $T_{cc}(3875)$ to the $D^0 D^{*+}$ and $D^+ D^{*0}$ thresholds, together with a moderate attractive interaction, will lead to a molecular state of these components \cite{25}, one cannot rule out that there could be a mixture, even if small, of a genuine compact tetraquark structure, as suggested in \cite{Yan:2021wdl}. The issue of the compositeness and elementariness of states has drawn much attention, starting by the well known work of Weinberg \cite{weinberg} supporting the deuteron as a bound state of a proton and a neutron. Improvements on the original idea of \cite{weinberg} have been done in \cite{migueljuan,2a,3a,4a,5a,6a,7a,8a,9a,10a,11a,12a,13a,14a,15a,16a,17a,18a,19a,20a,21a,hyodorev,daisong}.
 In \cite{daisong} the effect of the finite range of the interaction, implicitly assumed to be of zero range in coordinate space in \cite{weinberg}, is investigated and found to be relevant in the determination of the compositeness.

   For bound states the probability to have a certain component in the wave function  is given by $-g^2 \partial G/\partial s$, where $g$ is the coupling of the state to the given component and $G$ the loop function of the propagators of the two hadrons of the component, and for a potential that is energy independent the sum rule $\sum  -g_i^2 \partial G_i/ \partial s = 1$ holds, indicating that the state is fully molecular and the sum of probabilities of the different components is unity \cite{7a,17a,hyodorev}. However, if the potential is energy dependent, the former sum rule can be smaller than 1, accounting for the possibility of having other components in the wave function. In fact, it is possible to eliminate some channel resulting in an energy dependent effective potential for the remaining channels, and the sum rule of these latter channels gives unity minus the probability of the eliminated channel \cite{acetidai,8a,13a,hyodorev}. Similarly, the existence of a contribution from a genuine, or compact tetraquark in this case, can also be taken into account eliminating this channel in favor of an energy effective potential for the remaining channels \cite{sasa}.

   The works assuming a molecular picture for the $T_{cc}(3875)$ ordinarily assume energy independent interaction potentials between the components of the molecule, and hence, the sum rule is satisfied and the sum of probabilities of the $D^0 D^{*+}$ and $D^+ D^{*0}$ channels is unity. In the case of \cite{juanguo}, the probabilities of these states are 71\% and 29\% respectively \footnote{Actually in Ref. \cite{juanguo} there is a weak energy dependence in the potential stemming from an extra pion exchange mechanism in the interaction, in spite of which the state is still largely molecular.}, or the case of \cite{Cheng:2022qcm}, where the probabilities are also of the order of 72\% and 27\% respectively, or in \cite{feijooliang} with 69\% and 31\% \footnote{In \cite{feijooliang} the potentials, stemming from the local hidden gauge approach, have a weak energy dependence which however has a negligible influence in the results of the sum rule}, or in \cite{migueltcc} where the probabilities are 80\% and 20\%. There is a similarity in all these results, and the probability of the first channel is bigger than the second channel because its threshold is closer to the energy of the state.  This difference has not to be interpreted as a large violation of isospin. Indeed the $T_{cc}(3875)$ is assumed to have isospin $I=0$ and no trace of a peak is seen in the partner isospin $I=1$  channel $D^+ D^{*+}$. Isospin is a symmetry of strong interactions, which are of short range, and what matters in reactions where isospin is evidenced is the wave function at the origin (for s-wave), $g \,G$ according to  \cite{7a}. With values of $G$ similar for the two channels here, the couplings are what determine the isospin nature of the states. From \cite{feijooliang} we obtain values of $g_i$ for each channel which are of opposite sign and equal in size within 2-3\%, as it corresponds to an $I=0$ state. Since the probability is given by $-g^2 \partial G/ \partial s$ and $\partial G/ \partial s$ goes to infinity as we approach the threshold of the channel, it comes natural that, even having good isospin, the probability of the channel whose threshold is closer to the mass of the state will be larger than the one of the other channel. This is also what happens with the $D^0 \bar{D}^{*0}$ and $D^+ D^{*-}$ components in the $X(3872)$ \cite{daniiso}. The other thing to consider is that what determines the molecular nature of the state is the sum of the probabilities of the two channels that form the state.

   Very recently, two papers have addressed the issue of the probability of the
 $D^0 D^{*+}$ and $D^+ D^{*0}$ components.  One of them is \cite{misha}, where only the probability of the first channel is evaluated with a result of about 50-80\%.  Note that in this case the Weinberg prescription is used, which according to many posterior works, has been shown to be inaccurate in many cases \footnote{It suffices to mention that this prescription gives a molecular probability for the deuteron of 1.68 and the effective range, $r_0$, has opposite sign to the experimental one (see details in \cite{daisong}). However, as shown in \cite{4a}, the prescription works better when the effective range is negative.}. The other work is \cite{hyodocompo}, where a general discussion is done based on the value of the binding energy alone, together with several other assumptions made to validate the results. As we shall see, we will base our conclusions on two types of fits to the data, one fitting the binding, scattering lengths and effective ranges, seven data in our case, and another one making a fit to the $D^0 D^0 \pi^+$  distribution itself. This information obviously contains more dynamical information to make conclusions than the binding energy alone, which is clearly insufficient, since one can easily have a genuine state with a given binding energy, provided there is no interaction between the meson-meson components, or a pure molecular state from the interaction of the meson-meson components, provided there is no contribution from a genuine state, or such a state is very far away from the  mass of the observed state (see further discussions along this line in \cite{juan}.
We must also stress that in \cite{hyodocompo} the effect of the interaction range is also investigated  and it is found that, in addition to the binding energy of the $T_{cc}$, if the range of the interaction in momentum space is assumed  to be of the order of  $770\mev$, as it would correspond to vector exchange, the  compositeness of the $T_{cc}$ is much bigger than assuming a range due to pion exchange, and becomes of the order of 94\%.

    The former discussion shows that the subject is ripe to make a thorough discussion of the issue, which is our purpose here. For this purpose we shall include a general potential for the interaction of the  $D^0 D^{*+}$ and $D^+ D^{*0}$ channels, including the necessary terms with energy dependence to account for possible nonmolecular components, and carry fits to data, from where we will evaluate the probabilities of the two channels and $Z$, the probability of other possible nonmolecular components.  We shall see that the results strongly support the molecular picture, with a negligible probability of a nonmolecular component.

\section{Formalism}
The $T_{cc}(3875)$ is found according to Ref. \cite{expe2} with mass
\begin{eqnarray}
M_{T_{cc}}=M_{D^0 D^{*+}}+\delta m_{\rm exp}\,,
\end{eqnarray}
with $M_{D^0 D^{*+}}=3875.09\mev$ and
\begin{eqnarray}
\label{eq:mpole}
\delta m_{\rm exp} &=& -360\pm 40^{+4}_{-0}\kev \,,\nonumber
\end{eqnarray}
 and width
 \begin{eqnarray}
\Gamma &=& 48 \pm 2^{+0}_{-14}\kev \,.
\end{eqnarray}

The state is in $J^P=1^+$ and is assumed to be in isospin $I=0$, since no signal is seen in the $I=1$, $D^+ D^{*+}$ channel. The two channels
close to the $T_{cc}$ mass are $D^0 D^{*+}$ and $D^+ D^{*0}$, which we consider explicitly in our study. We have the
isospin doublets $(D^+,-D^0)$,$(D^{*+},-D^{*0})$ and hence
\begin{eqnarray}\label{eq:iso}
  |D^* D, I=0\rangle &=& -\frac{1}{\sqrt{2}} (D^{*+} D^0 - D^{*0} D^+)\,, \nonumber\\[0.10cm]
  |D^* D, I=1\rangle &=&- \frac{1}{\sqrt{2}} (D^{*+} D^0 + D^{*0} D^+)\,.
\end{eqnarray}

In our study we shall not assume a priori that the $T_{cc}$ is a state of $I=0$, it will come out from the analysis
of the experimental data. We shall only assume that isospin is a good symmetry for the interaction such that  the potential
 can be diagonalized  with the isospin basis. The different masses of $D^0 D^{*+}$ and $D^+ D^{*0}$, that go in the kinetic energy
 of the Hamiltonian, can, and will, produce a small violation of isospin in the final wave function.  This said, we take a general potential
 for the interaction of the two channels given by
\begin{eqnarray}\label{eq:v}
V = \left(
           \begin{array}{cc}
            V_{11} & V_{12}   \\[0.1cm]
           V_{12} &  V_{22}\\
           \end{array}
         \right)\,,
\end{eqnarray}
from where the scattering matrix is
\begin{eqnarray}\label{eq:BSeq}
T=[1-VG]^{-1} \, V,
\end{eqnarray}
with $G={\rm diag}(G_1, G_2)$, where $G_i$ are the loop functions for  the two channels which we
regularize in the cutoff method, with
\begin{eqnarray}\label{eq:Gcut}
G = \int_{|{\bm q}|<q_{\rm max}} \frac{d^3q}{(2\pi)^3} \, \frac{\omega_1 + \omega_2}{2 \,\omega_1  \omega_2} \,\frac{1}{s-(\omega_1 + \omega_2)^2+i\epsilon}
\end{eqnarray}
where $\omega_i = \sqrt{{\bm{q}}^2 +m_i^2}$, $m_1$ is the mass of the $D$ and $m_2$ that of $D^*$. The value of  $q_{\rm max}$
reflects the range of the interaction in momentum space \cite{7a,daisong} and will be obtained from the fits to the data.
Since we assume isospin symmetry for the potential we shall have
\begin{eqnarray}
\langle I=0 | V| I=1\rangle=0 \,,\nonumber
\end{eqnarray}
which gives us
\begin{eqnarray}\label{eq:x7}
V_{11}=V_{22}\,.
\end{eqnarray}
We also find that
\begin{eqnarray}\label{eq:ii}
\langle I=0 | V | I=0\rangle &=& V_{11}-V_{12} \,,\nonumber\\
\langle I=1 | V | I=1\rangle &=& V_{11}+V_{12}\,.
\end{eqnarray}

According to the findings of \cite{expe2} we should expect that  $V_{11}-V_{12}$ is negative, to allow for a bound state,
and  $V_{11}+V_{12}$ should be positive to prevent a bound state, from where we should have
\begin{eqnarray}\label{eq:con}
V_{12}>0\,;\qquad V_{12} >| V_{11}| \,.
\end{eqnarray}

However, we could relax these conditions, since, as we shall see, the state that we obtain has $I=0$ and all we need is that $V_{11}-V_{12}$ is
negative and $V_{11}+V_{12}$ relatively smaller in size than $V_{11}-V_{12}$, such that it  cannot lead to any bound state.

Then we have three parameters in the formalism, $V_{11}$, $V_{12}$ and $q_{\rm max}$. But to take into account the possibility
of a genuine component (compact tetraquark) or contribution from other meson-meson channels different than the two explicitly
considered, we will add terms in the potential which are energy dependent. According to the discussion in \cite{daisong,sasa} it is a
good approximation to retain just the terms linear in $s$. Thus we write
\begin{eqnarray}\label{eq:vv}
V_{11} &=& V'_{11} + \frac{\alpha}{m^2_V} (s-s_0) \,, \nonumber\\
V_{12} &=& V'_{12} + \frac{\beta}{m^2_V} (s-s_0)\,,
\end{eqnarray}
where $\alpha$, $\beta$ are free parameters, $m_V=800\mev$ is a typical vector mass, chosen to have  $\alpha$, $\beta$  dimensionless,
and $s_0$ is the mass squared of the $T_{cc}$. Our formalism has now five free parameters, but demanding that we have a pole at $s_0$, they
will be reduced to four. For the experimental data we will follow two different strategies:

\begin{itemize}
  \item[(a)]a fit to the $a$ and $r_0$ parameters
of the effective range expansion for the two channels,  $D^0 D^{*+}$ and $D^+ D^{*0}$, obtained from the experimental analysis  of the
data in  Ref. \cite{expe2};
  \item[(b)]a fit to the unitary amplitudes extracted  in the analysis of the experimental data in Ref. \cite{expe2}.
\end{itemize}

The scattering length  $a$ and effective range $r_0$ correspond to the effective range expansion of the scattering matrix
\begin{eqnarray}\label{eq:x1}
f(s)=\frac{1}{-\frac{1}{a}+\frac{1}{2} r_0 k^2-ik}\,.
\end{eqnarray}
Let us also mention here that in our formalism  with Eqs.~\eqref{eq:BSeq}, \eqref{eq:Gcut} for the scattering  amplitude we have
the relationship
\begin{eqnarray}\label{eq:x2}
T(s)= -8 \pi \sqrt{s} f(s)
\end{eqnarray}
which one can easily induce by looking at the imaginary parts of ${\rm Im}\, T^{-1}(s)$ = $-{\rm Im}\,G$ (for a real potential)
and $f^{-1}(s)$.

To avoid unnecessary complications due to the small width of the $D^*$, in the strategy (a) we shall disregard
the $D^*$ width \footnote{We thank Mikhail Mikhasenko for providing us the data in that limit.
We also recall that the prescription in Ref. \cite{expe2} has $a$ with opposite sign, which we change to our convention here.}. In the direct fit to the unitary amplitude of \cite{expe2} we shall take into account the $D^*$ width.
The two fits are  complementary and for fit (a)  we rely on the fact that the very small width of the $D^*$ will not change the character
of the  $T_{cc}$ state. Actually, we are used to call bound state to atoms and nuclei with unstable particles.  For strategy (a) we have
$a_1,r_{0,1}$ real for the  $D^0 D^{*+}$ channel and $a_2,r_{0,2}$ complex for the $D^+ D^{*0}$ channel. Hence, we have $6$ data to be
fitted with $4$ free parameters, for which we minimize the $\chi^2$ in the fit. The data of  \cite{expe2} neglecting the
 $D^*$ width are
\begin{eqnarray}\label{eq:mik}
a_1 &=& 6.134 \pm 0.51 ~\fm \,,  \nonumber\\[0.1cm]
r_{0,1} &=& -3.516 \pm 0.50 ~\fm  \,,\nonumber\\[0.1cm]
a_2 &=& (1.707 \pm 0.30) -i\, (1.07 \pm 0.30) ~\fm \,,\nonumber\\[0.1cm]
r_{0,2} &=& (0.259 \pm 0.30) -i\, (3.769 \pm 0.30)~ \fm \,.
\end{eqnarray}
The errors for $a_1, r_{0,1}$ are taken from \cite{expe2}, and for  $a_2, r_{0,2}$
are taken relatively of the same order of magnitude. The errors influence the value of $\chi^2$ but not the
parameters of the minimum of $\chi^2$ when this reaches values close to zero, as is the case here.

It is interesting to compare the results of Eq.~\eqref{eq:mik} with those induced from \cite{feijooliang}
\begin{eqnarray}\label{eq:feijoo}
 a_1 &=& 6.42~\fm \,,  \nonumber\\[0.1cm]
 r_{0,1} &=& -3.49~\fm  \,,\nonumber\\
 a_2 &=& (2.08 -i\, 0.98)~\fm \,, \nonumber\\[0.1cm]
 r_{0,2} &=& (-0.06-i\, 2.27)~\fm \,.
\end{eqnarray}
The results are very similar although not equal. The real parts of $r_{0,2}$ are different and even with opposite sign, but
let us note that they are one order of magnitude  smaller than their corresponding imaginary parts, and not incompatible
within the assumed uncertainties. Note that   $a_2$, $r_{0,2}$ are complex,  even neglecting the $D^*$ width because
the decay of $D^+ D^{*0}$ to $D^0 D^{*+}$ is open at the $D^+ D^{*0}$ threshold. The magnitudes of Eq.~\eqref{eq:feijoo} are
obtained with a potential
\begin{widetext}
\begin{eqnarray}\label{eq:prd}
V = g^2\, \frac{1}{2} \left[3s-(M^2+m^2+M'^2+m'^2) -\frac{1}{s}(M^2-m^2)(M'^2-m'^2) \right]
\left(
           \begin{array}{cc}
           \frac{1}{M_{J/\psi}^2}  & \frac{1}{m_\rho^2}     \\[0.1cm]
          \frac{1}{m_\rho^2}   &   \frac{1}{M_{J/\psi}^2}\\
           \end{array}
         \right) \,,
\end{eqnarray}
\end{widetext}
where $M, M'$ are the initial, final  vector masses and $m, m'$ the initial, final pseudoscalar masses, with $s$ calculated
at the threshold of  $D^0 D^{*+}$ and average masses of $D^0, D^+$ and $D^{*0}, D^{*+}$. The actual masses are used in the evaluation
of the scattering matrices of Eqs.~\eqref{eq:BSeq}, \eqref{eq:Gcut}. When we fit the magnitudes of  Eq.~\eqref{eq:feijoo} with the general potential of Eqs.~\eqref{eq:v}, \eqref{eq:vv} and the $T$ matrix of Eq.~\eqref{eq:BSeq}, we obtain a best fit with $\chi^2=0$ with $\alpha=\beta=0$ ,
certainly the potential of Eq.~\eqref{eq:prd}, and the sum rule
\begin{eqnarray}\label{eq:co1}
-\big(g^2_1 \,\frac{\partial G_1}{\partial s} + g^2_2 \,\frac{\partial G_2}{\partial s} \big)\big|_{s=s_0}  =1\,,
\end{eqnarray}
is exactly fulfilled indicating that we have a pure molecular state. This is actually a tautology, since we start with a potential
that produces a molecule and we get the answer that it is indeed a molecule. However, the test tells us that the inverse problem
of getting the potential parameters from the magnitudes $a_1, r_{0,1}, a_2, r_{0,2}$ is feasible and meaningful.
The interesting test is to obtain the potential of Eqs.~\eqref{eq:v}, \eqref{eq:vv} from the experimental data of Eq.~\eqref{eq:mik} where no assumptions concerning the interaction have been done, and then evaluate the probabilities
$P_1, P_2$ for the  $D^0 D^{*+}$ and $D^+ D^{*0}$  channels.\\

\section{Evaluation of the scattering matrix, the couplings and the probabilities of the channels}
\label{sec:3}

The evaluation of the $T$ matrix of Eq.~\eqref{eq:v} with two channels is easy and we show the formula below

\begin{widetext}
\begin{eqnarray}
T=\frac{1}{\rm DET}  \left(
           \begin{array}{cc}
            V_{11}+ ( V^2_{12}- V^2_{11}) G_2 & V_{12}   \\[0.1cm]
           V_{12} &  V_{11}+ ( V^2_{12}- V^2_{11}) G_1\\
           \end{array}
         \right)\,,
\end{eqnarray}
\end{widetext}
with the determinant of $(1-V G)$, DET, given by
\begin{eqnarray}
{\rm DET}=1- V_{11} (G_1+G_2)-( V^2_{12}- V^2_{11}) \,G_1 G_2 \,.
\end{eqnarray}

The bound state appears when ${\rm DET}=0$ at $s_0$, hence
\begin{eqnarray}
{\rm DET} \, (s=s_0)=0\,,
\end{eqnarray}
from where we can eliminate $V_{12}$ in terms of $V_{11}$, as (note that
the $\alpha,\beta$ terms of Eq.~\eqref{eq:vv} vanish at $s=s_0$)
\begin{eqnarray}\label{eq:v12p}
{V'}^2_{12}= \frac{1}{G_1 G_2} \left\{1-V'_{11}(G_1+G_2)+  {V'}^2_{11} G_1 G_2 \right\} \big|_{s=s_0}\,,
\end{eqnarray}
and we take the positive  root of $V_{12}$ according to Eq.~\eqref{eq:con}. The couplings
are defined from the residues of the $T$ matrix. Since
\begin{eqnarray}
T_{ij} \simeq \frac{g_i g_j}{s-s_0} \,,
\end{eqnarray}
then
\begin{eqnarray}\label{eq:coup}
g^2_1 &=& \lim_{s \to s_0} (s-s_0) \,T_{11}= \frac{V_{11}+(V^2_{12}- V^2_{11}) \, G_2 }{\frac{\partial}{\partial s} \, {\rm DET}} \big|_{s=s_0} \,,\nonumber\\
g^2_2 &=& \lim_{s \to s_0} (s-s_0) \,T_{22}= \frac{V_{11}+(V^2_{12}- V^2_{11}) \, G_1 }{\frac{\partial}{\partial s} \, {\rm DET}} \big|_{s=s_0} \,,\nonumber\\
g_1 g_2 &=& \lim_{s \to s_0} (s-s_0) \,T_{12}= \frac{V_{12}}{\frac{\partial}{\partial s} \, {\rm DET}} \big|_{s=s_0} \,,
\end{eqnarray}
where in the second step in the former equations we have used L'H\^{o}pital rule. Note that the last equation allows us to get the relative sign of $g_1$ and $g_2$. Then, according to \cite{7a,hyodorev} we have the probabilities for the $D^0 D^{*+}$ and $D^+ D^{*0}$ channels, respectively, as
\begin{eqnarray}
P_1 = -g^2_1\, \frac{\partial G_1}{\partial s}\big|_{s=s_0} \,,\quad  P_2 = - g^2_2 \,\frac{\partial G_2}{\partial s}\big|_{s=s_0} \,,
\end{eqnarray}
and consequently, the nonmolecular part, that could also account for possible missing coupled channels, is
\begin{eqnarray}
Z=1-P_1-P_2\,.
\end{eqnarray}

With the explicit formulae for the couplings, Eq.~\eqref{eq:coup}, it is  now easy to prove explicitly
the sum rule of  Eq.~\eqref{eq:co1} when $\alpha=\beta=0$, although a general proof, extended to any
number of channels, can be found in   \cite{7a,hyodorev}, identifying each term of the sum rule
with the probability of the corresponding channel in the wave function of the state.

\section{Evaluation of the scattering lengths  and effective ranges  }

From Eqs.~\eqref{eq:x1}, \eqref{eq:x2} we find
\begin{eqnarray} \label{eq:im}
-\frac{1}{a}+\frac{1}{2} r_0 \, k^2 - i \,k \equiv  -8 \pi \, \sqrt{s} \, T^{-1} \,.
\end{eqnarray}
Hence, for each of the two channels $D^0 D^{*+}$ and $D^+ D^{*0}$, we have
 \begin{eqnarray}
 {\rm Im} \, T^{-1}_{ii} = - {\rm Im} \,G_i =\frac{1}{8 \pi \, \sqrt{s}} k_i  \,, \nonumber
\end{eqnarray}
with

\begin{eqnarray}
k_i=\frac{\lambda^{1/2}(s,M^2_i,m^2_i)}{2\, \sqrt{s}} \,,\nonumber
\end{eqnarray}
and then we can  write

\begin{widetext}
\begin{eqnarray}
-\frac{1}{a_1}&=& -8 \pi \, \sqrt{s}  \,T^{-1}_{11}  = -8 \pi \, \sqrt{s}   \left[\frac{1-V_{11}G_2}{V_{11}+(V^2_{12}-V^2_{11}) \,G_2}
-Re\,G_1\right]\big|_{s=s_1}  \,,\nonumber\\
r_{0,1}&=& -\frac{\sqrt{s_1}}{\mu_1} \frac{\partial}{\partial s} \left\{16 \pi \, \sqrt{s} \left[\frac{1-V_{11}G_2}{V_{11}+(V^2_{12}-V^2_{11}) \,G_2}-Re \, G_1\right]\right\}\big|_{s=s_1}\,,
\end{eqnarray}
\end{widetext}
with $s_i$ the square of the threshold energies for the two channels and where to calculate $r_{0,1}$ we have used that
 \begin{eqnarray}
 \frac{\partial}{\partial k^2}= \frac{\partial s}{\partial k^2} \frac{\partial}{\partial s} =\frac{\sqrt{s}}{\mu} \frac{\partial}{\partial s} \,,\nonumber
\end{eqnarray}
with  $\mu$ the reduced mass of $M,m$
\begin{eqnarray}
\mu=\frac{M m}{M+m} \,.\nonumber
\end{eqnarray}

Similarly,
\begin{widetext}
\begin{eqnarray}
-\frac{1}{a_2}&=& -8 \pi \, \sqrt{s}  \,T^{-1}_{22}  = -8 \pi \, \sqrt{s}   \left[\frac{1-V_{11}G_1}{V_{11}+(V^2_{12}-V^2_{11}) \,G_1}
-Re\, G_2\right]\big|_{s=s_2} \,, \nonumber\\
r_{0,2}&=& -\frac{\sqrt{s_2}}{\mu_2} \frac{\partial}{\partial s} \left\{16 \pi \, \sqrt{s} \left[\frac{1-V_{11} G_1}{V_{11}+(V^2_{12}-V^2_{11})\, G_1}
-Re\, G_2\right]\right\}\big|_{s=s_2}\,.
\end{eqnarray}
\end{widetext}

\section{Results}

\subsection{Fit to the data of $a$ and $r_0$ }

We first conduct the fit (a) to the data of $a, r_0$ of Mikhasenko \cite{misha} given in Eq.~\eqref{eq:mik}. We define
\begin{eqnarray}\label{eq:chisq}
\chi^2=\sum_i \left(\frac{y^{th}_i-y^{exp}_i}{\Delta y^{exp}_i}\right)^2 \,,
\end{eqnarray}
where $y^{exp}_i$ are $a_1,r_{0,1}$  which are real, and  $a_2,r_{0,2}$ which are complex. In the end we have six data (the binding
energy, was used in Eq.~\eqref{eq:v12p} to eliminate $V'_{12}$), and four parameters $V'_{11}, \alpha, \beta, q_{\rm max}$.
The reduced $\chi^2_r$ is $\chi^2/2$. We minimize  $\chi^2_r$ to obtain these parameters and, once  obtained, we calculate
$P_1, P_2, Z$. Given the fact that there are correlations between the parameters, as we shall see,
we find useful in order to find the uncertainties in the output to use a method often  used in these works \cite{54b,55b,56b,57b,58b,59b,60b,61b}.
This consist in calculating  $\chi^2$ of Eq.~\eqref{eq:chisq}, using instead of $y^{exp}_i$ a random number within
the interval $[y^{exp}_i-\Delta y^{exp}_i, y^{exp}_i+\Delta y^{exp}_i]$. We conduct several runs
in this way, selecting the good ones with $\chi^2_r <0.4$, and for each run we determine the magnitudes  $P_1, P_2, Z$
(and also  any other magnitude that one wished to obtain). The errors obtained for the different magnitudes in this way do not change if we
take $\chi^2_r < 0.4,0.3,0.2,0.1$ and the average values are always within these error bars.
Then, we evaluate the average value of these magnitudes and the dispersion in the usual way, for instance
\begin{eqnarray}
\overline{P}_1 &=&\frac{1}{N} \sum_i {P}_{1,i} \,,\nonumber \\
 (\Delta {P}_1)^2  &=&  \frac{1}{N} \sum_i ({P}_{1,i}-\overline{P}_1)^2 \,.
\end{eqnarray}

\begin{table*}[ht!]
\centering
\renewcommand\arraystretch{1.5}
\caption{\label{tab:dai} The obtained scattering lengths and effective ranges. }
\begin{tabular}{cccccc}
\hline\hline
 & $a_1~[\fm]$ & $r_{0,1}~[\fm]$ &  $a_2~[\fm]$ & $r_{0,2}~[\fm]$ & \\
\hline
 & $6.110 \pm 0.065$ & $-3.455 \pm 0.194$ & $ (1.761\pm 0.031) -i \, (1.063 \pm 0.024)$ & $(0.265 \pm 0.148)-i\, (3.760 \pm 0.142)$  &\\
\hline\hline
\end{tabular}
\end{table*}

The results that we obtain are shown in Table \ref{tab:dai}.
The agreement of the values obtained for the scattering lengths and effective
range with the experimental ones of Eq.~\eqref{eq:mik} is remarkable.

\begin{table*}
\centering
\renewcommand\arraystretch{1.5}
\caption{\label{tab:daig} The obtained coupling constants and probabilities.}
\begin{tabular}{cccccccc}
\hline\hline
 &  $g_1$ [MeV] & $g_2$ [MeV] &  $P_1$ & $P_2$ &  $Z$\\
\hline
& $3759.88 \pm 32.95$  & $-3820.52 \pm 43.61 $  &  $0.685 \pm 0.011$ & $0.285 \pm 0.005$ & $0.030\pm 0.016$ \\
\hline\hline
\end{tabular}
\end{table*}

As we see in Table \ref{tab:daig},  we obtain $P_1, P_2$ of the order of 69\%, 29\% with uncertainties of the order of 1\%,
and $Z=0.03 \pm 0.02$. The numbers obtained qualify the $T_{cc}$ as a clear molecular state made of the  $D^0 D^{*+}$ and $D^+ D^{*0}$
components. We also find there that the couplings $g_1,g_2$ are very close
to each other and of opposite sign, indicating, according to Eq.~\eqref{eq:iso} that we have basically a state of $I=0$.
It is illustrating to take one typical fit of those obtained. We find
\begin{widetext}
\begin{eqnarray}\label{eq:daifit}
q_{\rm max}=700\mev, \quad V'_{11}=-100,\quad  V'_{12}=146.07, \quad \alpha=-546.6, \quad \beta=-498.25 \,.
\end{eqnarray}
\end{widetext}
This contrasts with the fit obtained in \cite{feijooliang} using the potential of Eq.~\eqref{eq:prd} and $\alpha=\beta=0$,

\begin{widetext}
\begin{eqnarray}\label{eq:feijoofit}
q_{\rm max}=420\mev, \quad V'_{11}=28.37,\quad  {\rm V'_{12}}=459.87, \quad \alpha=0, \quad \beta=0 \,.
\end{eqnarray}
\end{widetext}

We should first realize that in a range of about $5\mev$ from the energy of the $T_{cc}$ the terms $\alpha (s-s_0)$,
$\beta (s-s_0)$ are of the order of 16\% of $V_{12}$ of Eq.~\eqref{eq:daifit}, or
 5\% compared to  $V_{12}$ of Eq.~\eqref{eq:feijoofit}, hence, a relatively small magnitude. Yet,
 it looks like the solutions of Eqs.~\eqref{eq:daifit}, \eqref{eq:feijoofit} are very different.
 Indeed, the values of $q_{\rm max}$, although qualitatively similar, are still different, and the strength of the
 potential is also rather different. However, this only reflects the strong correlations between the parameters.
 Indeed, if we look only to the binding energy (but we have more information) and assuming we just have the
 $I=0$ state that we have induced from the data, we would have in one channel
 \begin{eqnarray}
 T=[1-VG]^{-1}V=\frac{1}{V^{-1}-G} \,,
\end{eqnarray}
where, according to Eq.~\eqref{eq:ii}
\begin{eqnarray}
V=V_{11}-V_{12}\,.
\end{eqnarray}
In order to get the binding at $s_0$ we have  $V^{-1}=G(s_0)$. Hence there is a trade off between $V$
and $G$, such that making changes with $\delta V^{-1}=\delta G$ we would get the same binding. Then
starting with $V$ from \cite{feijooliang}, let us call it $V_{\rm FL}$, and $G$ with $q_{\rm max}=420\mev$, we
would get an equivalent $V_{\rm EQ}$ for $q_{\rm max}=700\mev$ such that
\begin{eqnarray}
V_{\rm EQ}^{-1}-V_{\rm FL}^{-1}=G(q_{\rm max}=700\mev)-G(q_{\rm max}=420\mev) \,. \nonumber
\end{eqnarray}
With this we obtain
\begin{eqnarray}
V_{\rm EQ}=-245.27  \,,
\end{eqnarray}
to be compared to $V$ from Eq.~\eqref{eq:daifit}, let us call it  $V_{\rm Mi}=V'_{11}-V'_{12}$
\begin{eqnarray}
V_{\rm Mi}=-246.07 \,.
\end{eqnarray}

This is also an extra information obtained from the data.
It is also interesting to note that  we find many fits with $\chi^2_r$ smaller than $0.3$ with
different values of $\alpha$ and $\beta$. Yet, one can see that there is a strong correlation
between the values of $\alpha$ and $\beta$, always very similar in size and such that $(\alpha-\beta)(s-s_0)/m^2_V$, as it
corresponds to the $I=0$ state that we find, is very small, always smaller  than 1\% of $V'_{12}$ of Eq.~\eqref{eq:feijoofit} in the range
of $5\mev$ above the $\sqrt{s_0}$. The smallness of this number is pointing to the tiny nonmolecular component of the $I=0$ state.

The agreement found is remarkable, stressing once more the $I=0$ character of the $T_{cc}$ state.
Considering the probabilities  $P_1, P_2$ we observe that what we obtain in Table \ref{tab:dai},
is essentially the same result as was found from the analysis of \cite{feijooliang} is spite of the apparent
different solutions in the  fitting parameters. This simply indicates strong correlations between
the parameters, yet within a range of natural values, like having the range $q_{\rm max}$ between
$420\mev$ and $700\mev$, typical of the exchange of a light vector meson. This is also very valuable information.

\subsection{Direct fit to the $D^0 D^0 \pi^+$ mass distribution}

Now we turn to fit (b). For it we take the $D^0 D^0 \pi^+$ mass distribution obtained in Ref. \cite{expe2} which corrects the
raw data by the experimental resolution and fits the distribution with a unitary amplitude,
accounting for the decay of the $D^*$. We compare then our results for $\Gamma (s)$ in Eqs. (18), (19) of Ref. \cite{feijooliang},
using the new potential of Eqs.~\eqref{eq:v}, \eqref{eq:x7}, \eqref{eq:vv}, with the results in Fig. 8 of the supplementary
information of Ref. \cite{expe2}. The parameters are $V'_{11}, V'_{12}, \alpha, \beta, q_{\rm max}$ and a normalization constant. There is a strong correlation between $V'_{11}, V'_{12}$ and $q_{\rm max}$. Indeed, if we use a non relativistic $G$ function \cite{7a} we observe that in one channel $t=(V^{-1}-G)^{-1}$ with $Re G \approx  -8\pi \mu (q_{\rm max} -\frac{k^2}{2q_{\rm max}}+O(k^4)) $ (with $\mu$ the reduced mass) and there would be relativistic corrections on the $k^2$ term. We see that what matters in $V^{-1}-G$ is $ V^{-1}+8\pi \mu q_{\rm max}$ at leading order and there is a trade off between $V$ and $q_{\rm max}$. By contrary, the effective range depends on $q_{\rm max}$, the range of the interaction, as shown in the works \cite{4a,hyodorev,daisong,Albaladejo:2022sux}. Since we are in a case very close to threshold, we wish to take advantage of this correlation and we will fix $q_{\rm max}$ to the one of Ref. \cite{feijooliang} ($q_{\rm max}=418.6$ MeV) and, in addition, we will also make fits with $q_{\rm max}=400, 550 \, {\rm and}  \, 700$ MeV to calculate what we will call systematic errors. Thus, we leave
$V'_{11}, V'_{12}$ as free parameters without using the constraint of Eq.~\eqref{eq:v12p} on the binding energy, since the mass distribution
contains information on the position of the peak.  On the other hand, we are using now the $G$ functions accounting for the width of the $D^*$ as done in \cite{roca} with
\begin{widetext}
\begin{eqnarray}\label{eq:Gcut1}
G = \int_{|{\bm q}|<q_{\rm max}} \frac{d^3q}{(2\pi)^3} \, \frac{\omega_1 + \omega_2}{2 \,\omega_1  \omega_2} \,\frac{1}{\sqrt{s}+\omega_1 + \omega_2} \,\frac{1}{\sqrt{s}-\omega_1 - \omega_2 + \,i\, \frac{\sqrt{s'}}{2\,m_{D^*}} \Gamma_{D^*}(s')} \,,
\end{eqnarray}
\end{widetext}
where $s'=(\sqrt{s}-\omega_D)^2-{\bm q}^2$  and $\Gamma_{D^*}(s')$ as given in Eqs. (14), (15) of \cite{feijooliang}.

\begin{figure*}[h!]
\centering
\includegraphics[scale=.45]{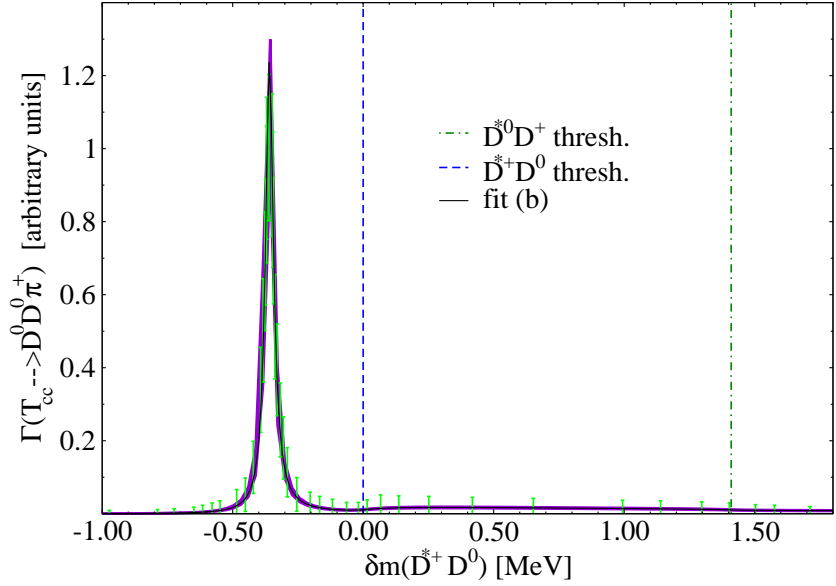}
\caption{Fit to the data of Fig. 8 of supplementary material in Ref. \cite{expe2} with the size of the errors taken from the raw data in the way explained in the text.}
\label{fig:new}
\end{figure*}

 We take $44$ points from Fig. 8 of Ref. \cite{expe2}, with the typical errors obtained from the raw data taking a $10\%$ value of the square root of the number of events, and perform a best fit to the data whose result is shown in Fig.~\ref{fig:new}. In order to estimate the statistical uncertainties we use the resampling (bootstrap) method \cite{54b,55b,56b,57b,58b,59b,60b,61b} and, as mentioned above, we will estimate the systematic errors for the observables from the use of different $q_{\rm max}$ values. The best fit returns the following parameters:
\begin{widetext}
\begin{eqnarray}\label{eq:feijfit}
\quad V'_{11}=-66 \pm 67, \quad  V'_{12}=367 \pm 67 , \quad \alpha=0 \pm 68\,, \quad \beta= 0\pm 55\,.
\end{eqnarray}
\end{widetext}
One should not worry too much about these values and their errors, since we know that there are also correlations between $V'_{11}$ and $ V'_{12}$ and $\alpha$ and $\beta$ (the differences between these parameters is what matters if one has an $I=0$ state). The relevant thing is what we get for the observables, and from the different fits in the resampling method we obtain the average values of the observables and their dispersion. The results are summarized in Tables \ref{tab:feijoo} and \ref{tab:feijoog}.

\begin{table*}[ht!]
\centering
\renewcommand\arraystretch{1.5}
\caption{\label{tab:feijoo} The obtained scattering lengths and effective ranges. The values between brackets correspond to the systematic error.}
\begin{tabular}{ccc}
\hline\hline
 & $a_i~[\fm]$ & $r_{0,i}~[\fm]$ \\
\hline
$i=1$ ($D^{*+}D^0$) & $ (7.60 \pm 0.13 (\pm 0.13) ) -i \, (1.73 \pm 0.06 (\pm 0.06))$ & $ -2.94 \pm 0.03 (\pm 0.27) $ \\
 $i=2$ ($D^{*0}D^+$)& $ (1.99\pm0.06 (\pm 0.06) ) -i \, (1.25 \pm 0.05 (\pm 0.06))$ & $( 0.11 \pm 0.12 (\pm 0.28) )-i\, (2.74 \pm 0.19 (\pm 0.21) )$  \\
\hline\hline
\end{tabular}
\end{table*}

\begin{table*}
\centering
\renewcommand\arraystretch{1.5}
\caption{\label{tab:feijoog} The obtained binding energy, width, coupling constants and probabilities. The values between brackets correspond to the systematic error.}
\begin{tabular}{cccccc}
\hline\hline
  $B$ [KeV] &  $\Gamma$ [KeV] &  $g_1$ [MeV] & $g_2$ [MeV] &  $P_1$ & $P_2$ \\
\hline
 $360 \pm 2(\pm 1) $  & $38 \pm 2 (\pm 1)  $ & $3875 \pm 50 (\pm 53)$  & $ -4077 \pm 63 (\pm 59)  $  &  $0.697 \pm 0.017 (\pm 0.008)$  &$ 0.301 \pm 0.009 (\pm 0.007)$ \\
\hline\hline
\end{tabular}
\end{table*}

These results are very similar to those displayed in Table \ref{tab:dai}.  Yet, they should be better compared to the results of Ref. \cite{expe2} when the $D^*$ width is explicitly taken into account, which are

\begin{eqnarray}\label{eq:exp}
a_1^{exp}=[(7.16 \pm 0.51)  -i \, (1.85 \pm 0.28)]~\fm  \,, \quad  a_2^{exp}=(1.76 -i \, 1.82)~\fm \,.
\end{eqnarray}

The agreement is perfect within errors if we assume the  relative errors  in $a_2^{exp}$ to be  similar to those of $a_1^{exp}$. We see that $a_1$ is now complex, yet with the imaginary part reasonably smaller than the real one. The real parts obtained for $a_1$ or $a_1^{exp}$  in this case are also very similar to $a_1$  of  Eqs.~\eqref{eq:mik} and \eqref{eq:feijoo} obtained with fit (a) to the data neglecting the width of the $D^*$.

We have taken advantage of the weak dependence of the obtained magnitudes  with the $D^*$ decay width and evaluate $P_1, P_2,P_1+P_2$ ($P_1+P_2=0.998 \pm 0.025(\pm 0.0004)$) \footnote{$P_1+P_2$ is calculated in each fit and the average of $P_1+P_2$ and its dispersion is obtained from the statistical analysis.} and $r_{0,1}$,  $r_{0,2}$, $g_1, g_2$ in the limit of $\Gamma_{D^*} \to 0$, using the parameters obtained in Eq.~\eqref{eq:feijfit} and the formulas described in section~\ref{sec:3}.

We see that the systematic errors in the effective ranges are bigger than for the scattering lengths, as one might expect, given the dependence of these variables on $q_{\rm max}$. The values of the couplings obtained are also very similar to those shown in Table \ref{tab:daig} with the fit (a), compatible within errors. Coming now to the compositeness $P_1,P_2$, the values obtained are also remarkably similar to those in Table \ref{tab:daig}, indicating again the molecular nature of the  $T_{cc}$ state with $P_1+P_2$ essentially $1$ in the present fit with an uncertainty of the order of $0.025$.

\section{limiting case of a nonmolecular state}

\begin{figure*}
\centering
\includegraphics[scale=.85]{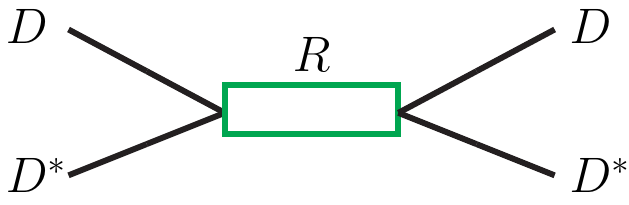}
\caption{Amplitude for   $D D^*$   from a preexisting nonmolecular  state.}
\label{fig:1}
\end{figure*}

\begin{figure*}
\centering
\includegraphics[scale=.85]{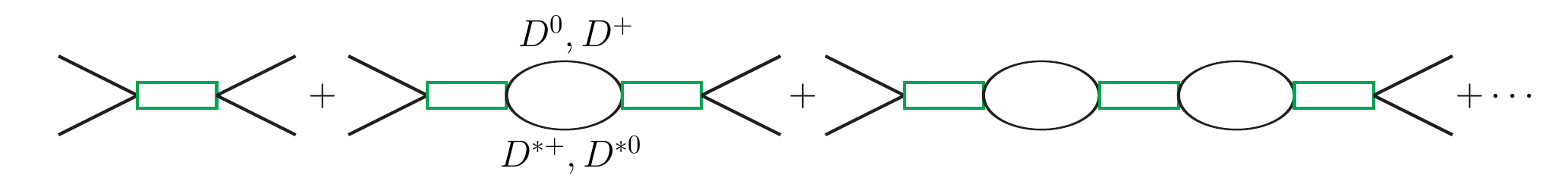}
\caption{Diagrammatic representation of the $T$ matrix for the nonmolecular  state including the $D D^*$ selfenergy. }
\label{fig:2}
\end{figure*}

We shall assume that the origin of the $T_{cc}$ state corresponds to a state, which has a very small overlap with the  $D^0 D^{*+}$, $D^+ D^{*0}$ components (a minimum overlap is needed such that the state is observed in the  $D^0 D^{*+}$ and $D^+ D^{*0}$ channels).  We shall also assume, to be consistent with experiment, that the state has $I=0$, although the conclusions are not tied to this fact. The amplitude for the $D D^*$ $(I=0)$ state will be given by
\begin{eqnarray}
\widetilde{T}_{D^*D,D^*D}=\frac{g^2}{s-s_0}  \nonumber
\end{eqnarray}
depicted in  Fig.~\ref{fig:1}, and $g^2$ should be small to prevent the overlap of the genuine state with the
$D D^*$ components. Consistently with the nonmolecular assumption, we will also assume that the
$D^0 D^{*+}$, $D^+ D^{*0}$  components have no interaction stemming from a potential,  or $t$-channel exchange of other hadrons.  Yet, the picture for the $T$ matrix is not complete since, for consistency with the small coupling of the genuine state to the $D D^*$   components, we  have to consider the selfenergy of the state due to its coupling to the $D D^*$   components, as depicted in   Fig.~\ref{fig:2}.

By taking into account the $I=0$ structure of the  $D D^*$  state, we find from  Fig.~\ref{fig:2}

\begin{widetext}
\begin{eqnarray}\label{eq:tx}
T_{D^0 D^{*+},D^0 D^{*+}}=\frac{1}{2} \, \frac{g^2}{s-\widetilde{s_0}-\frac{1}{2}g^2 G_{D^0 D^{*+}}-\frac{1}{2}g^2 G_{D^+ D^{*0}}}  \,.
\end{eqnarray}
\end{widetext}

Once again, by considering Eq.~\eqref{eq:im} and
that ${\rm Im}\,T^{-1} = \frac{1}{8 \pi \sqrt{s}} k$, we realize that the amplitude
of Eq.~\eqref{eq:tx} satisfies  unitarity in the two channels $D^0 D^{*+}$, $D^+ D^{*0}$, reproducing exactly the imaginary  part of $f^{-1}(s)$. We now derive the results for $a,r_0$ in the limit of $g \to 0$, such that  Eq.~\eqref{eq:tx} has a pole at $s_0$, where it should be.
Let us call D the denominator of Eq.~\eqref{eq:tx},
\begin{eqnarray}
{\rm D} =s-\widetilde{s_0} -\frac{1}{2}g^2 G_{D^0 D^{*+}}-\frac{1}{2}g^2 G_{D^+ D^{*0}} \,.
\end{eqnarray}
In order to have a pole of $T_{D^0 D^{*+},D^0 D^{*+}}$ at $s_0$, we need
\begin{eqnarray}
{\rm D}\big|_{s=s_0}=0  \,.
\end{eqnarray}


Let us note that if $g \neq 0$, then $\widetilde{s_0}\neq s_0$ and  we can make an expansion of $T$ in powers of $s-s_0$, and we would then
go back to the previous analysis accounting
for the genuine state in terms of the energy dependent terms, the constant parts reabsorbed in the
$V_{ij}$ coefficients. Thus,  the novel thing should be found in the $g^2 \to 0$ limit.

In this case it is easy to find $a$ and $r_0$ from  Eq.~\eqref{eq:tx}
\begin{widetext}
\begin{eqnarray}
-\frac{1}{a} = \lim_{g^2 \to 0} -8 \pi \sqrt{s} \, \frac{2}{g^2} \left\{s-s_0-\frac{1}{2}g^2 G_{D^0 D^{*+}}-\frac{1}{2}g^2 G_{D^+ D^{*0}}  \right\} \,.
\end{eqnarray}
\end{widetext}

Thus, we immediately  see that $-\frac{1}{a} \to \infty$ and hence $a \to 0$; $ a_1 \to 0$,  $a_2 \to 0$. \\

For the radius we find

\begin{widetext}
\begin{eqnarray}
\frac{1}{2} \, r_{0,1}  = \lim_{g^2 \to 0} -\frac{\sqrt{s_1}}{\mu_1} \,\frac{16\pi}{g^2} \,  \frac{\partial}{\partial s}\left\{s^{1/2}(s-s_0)-O(g^2)\right\}
=  \lim_{g^2 \to 0} -\frac{\sqrt{s_1}}{\mu_1} \,\frac{16\pi}{g^2} \,  \frac{1}{2 \sqrt{s}} (3s-s_0)\big|_{s=s_1} \, \to  -\infty  \,,\nonumber \\
 \frac{1}{2} \, r_{0,2}  =  \lim_{g^2 \to 0} -\frac{\sqrt{s_2}}{\mu_2} \,\frac{16\pi}{g^2} \,  \frac{1}{2 \sqrt{s}} (3s-s_0)\big|_{s=s_2} \, \to  -\infty \,.
\end{eqnarray}
\end{widetext}
We see that in the strict limit of the nonmolecular state, the scattering length goes to $0$ and the radius
to $ -\infty$. This contrast with the experimental results of Eq.~\eqref{eq:mik} where $a$ is relatively large for hadronic reactions, and $r_0$ is smaller than $a$.
A  discussion on the effects of the genuine state along the line described here, but not addressing the scattering length and
effective range can be seen in \cite{migueltcc} (see also section 2 of Ref. \cite{sasa}).

\section{Conclusions}
We have performed fits to the data of the LHCb collaboration on the $T_{cc}(3875)$ state with a general framework that contains explicitly the
 $D^0 D^{*+}$ and $D^+ D^{*0}$  components, which are allowed to interact with a potential containing energy dependent terms by means of which one can account for contributions of missing coupled channel, as well as effects from some genuine state, like a compact tetraquark state. We have conducted two types of fits to the data. One of them is a fit to data from LHCb for the scattering lengths and effective ranges of
 the channels $D^0 D^{*+}$ and $D^+ D^{*0}$, evaluated ignoring the $D^*$ width. The other fit is done directly to the data of the mass distribution of $D^0 D^0 \pi^+$ corrected by the experimental resolution and parametrized in terms of a unitary amplitude in \cite{expe2}, where the $D^*$ width is explicitly considered. In all fits we have calculated the parameters of the theoretical framework and from them we have looked at the bound state produced and evaluated the probabilities $P_1, P_2$ of  having  $D^0 D^{*+}$ and $D^+ D^{*0}$ in the wave function, as well as the couplings $g_1,g_2$ of the state obtained to these two components. Without any assumption of the  nature of the state observed, we find that the couplings obtained are very close to each other and of opposite sign, indicating the isospin $I=0$ nature of the state observed \footnote{Mathematically speaking the formulas for $T_{11}$ and $T_{22}$ depend only on $V_{12}^2$ and $T_{12}$ is proportional to $V_{12}$. This means that we could get solutions with $V_{12}$ positive or negative and eventually interprete the state as having $I=1$ (see discussion in \cite{migueltcc}). We reject the cases with positive $V_{11}-V_{12}$ based on the results of the local hidden gauge formalism \cite{feijooliang}, where $V_{11}-V_{12}$ is negative without any ambiguity (Eq. \ref{eq:prd}).}. The potential obtained produces a repulsive interaction in  $I=1$, justifying the non observance of any structure with $I=1$ in the LHCb experiment. On the other hand, from the couplings obtained, we could calculate the probabilities $P_1, P_2$ of $D^0 D^{*+}$, $D^+ D^{*0}$  and $Z=1-P_1-P_2$, the non compositeness of the state. We found values for $Z$ of the order of 3\% with an uncertainty that makes it compatible with zero or, conversely, in the case of fit (b) $P_1+P_2=1$ with uncertainties of 2.5\%, indicating that we have a very clear case of a molecular state made of the $D^0 D^{*+}$, $D^+ D^{*0}$ components. We found that the probabilities of the two channels are of the order of 69\% and 29\% respectively, very similar to what has been obtained in other approaches which, however, assume a priori energy independent potentials which do not allow nonmolecular components to appear. We also clarified that the isospin nature should not be induced by the probabilities $P_1,P_2$, but from the couplings $g_1,g_2$, or related to them, the wave functions at the origin of these two channels, since in strong interactions, where the isospin symmetry holds, the wave functions at short distances is what matters. We also shortly discussed that using all the available experimental information, apart from the mass of the state, was essential to reach the present conclusions, with the information on the binding being clearly insufficient to reach these conclusions.

\vskip .5cm
\section*{Acknowledgments}
We would like to thank M. Mikhasenko for useful discussions and for providing us with data of his analysis.
We also thank  M. Albaladejo  for useful discussions.
This work is partly  supported by the National Natural Science Foundation of China
under Grants Nos. 12175066, 11975009 and  LRD would like to express special thanks
to her students of Yuhang Wang and Langning Chen in helping to read the fitting data.
LMA has received funding from the Brazilian agencies Conselho Nacional
de Desenvolvimento Cient\'ifico e Tecnol\'ogico (CNPq) under contracts 309950/2020-1, 400215/2022-5, 200567/2022-5,
and Funda\c{c}\~ao de Amparo \`a Pesquisa do Estado da Bahia (FAPESB) under the contract INT0007/2016.
 This work is also partly supported by the Spanish Ministerio de
Economia y Competitividad (MINECO) and European FEDER funds under Contracts No. FIS2017-84038-C2-1-P
B, PID2020-112777GB-I00, and by Generalitat Valenciana under contract PROMETEO/2020/023. This project has
received funding from the European Union Horizon 2020 research and innovation programme under the program
H2020-INFRAIA-2018-1, grant agreement No. 824093 of the STRONG-2020 project.
The work of AF was partially supported by the Generalitat Valenciana and European Social Fund APOSTD-2021-112.


\end{document}